\documentclass[manuscript]{acmart}
\AtBeginDocument{%
  }

\setcopyright{acmlicensed}
\copyrightyear{2025}
\acmYear{2025}
\acmDOI{}
\acmConference[CI '25]{ACM Collective Intelligence}{August 04--06, 2025}{La Jolla, CA}

\usepackage{xcolor}

\begin{document}

\title{User-Centered Design with AI in the Loop: A Case Study of Rapid User Interface Prototyping with ``Vibe Coding''}



\author{Tianyi Li}
\email{li4251@purdue.edu}
\affiliation{%
  \institution{Purdue University}
  \city{West Lafayette}
  \state{Indiana}
  \country{USA}
}
\author{Tanay Maheshwari}
\email{mahesh88@purdue.edu}
\affiliation{%
  \institution{Purdue University}
  \city{West Lafayette}
  \state{Indiana}
  \country{USA}
}
\author{Alex Voelker}
\email{alex@voelker.org}
\affiliation{%
  \institution{Purdue University}
  \city{West Lafayette}
  \state{Indiana}
  \country{USA}
}






\renewcommand{\shortauthors}{Tianyi Li, Tanay Maheshwari, and Alex Voelker}

\begin{abstract}
We present a case study of using generative user interfaces, or ``vibe coding,'' a method leveraging large language models (LLMs) for generating code via natural language prompts, to support rapid prototyping in user-centered design (UCD). Extending traditional UCD practices, we propose an AI-in-the-loop ideate-prototyping process. We share insights from an empirical experience integrating this process to develop an interactive data analytics interface for highway traffic engineers to effectively retrieve and analyze historical traffic data. With generative UIs, the team was able to elicit rich user feedback and test multiple alternative design ideas from user evaluation interviews and real-time collaborative sessions with domain experts. We discuss the advantages and pitfalls of vibe coding for bridging the gaps between design expertise and domain-specific expertise.
\end{abstract}


\begin{teaserfigure}
  \includegraphics[width=\textwidth]{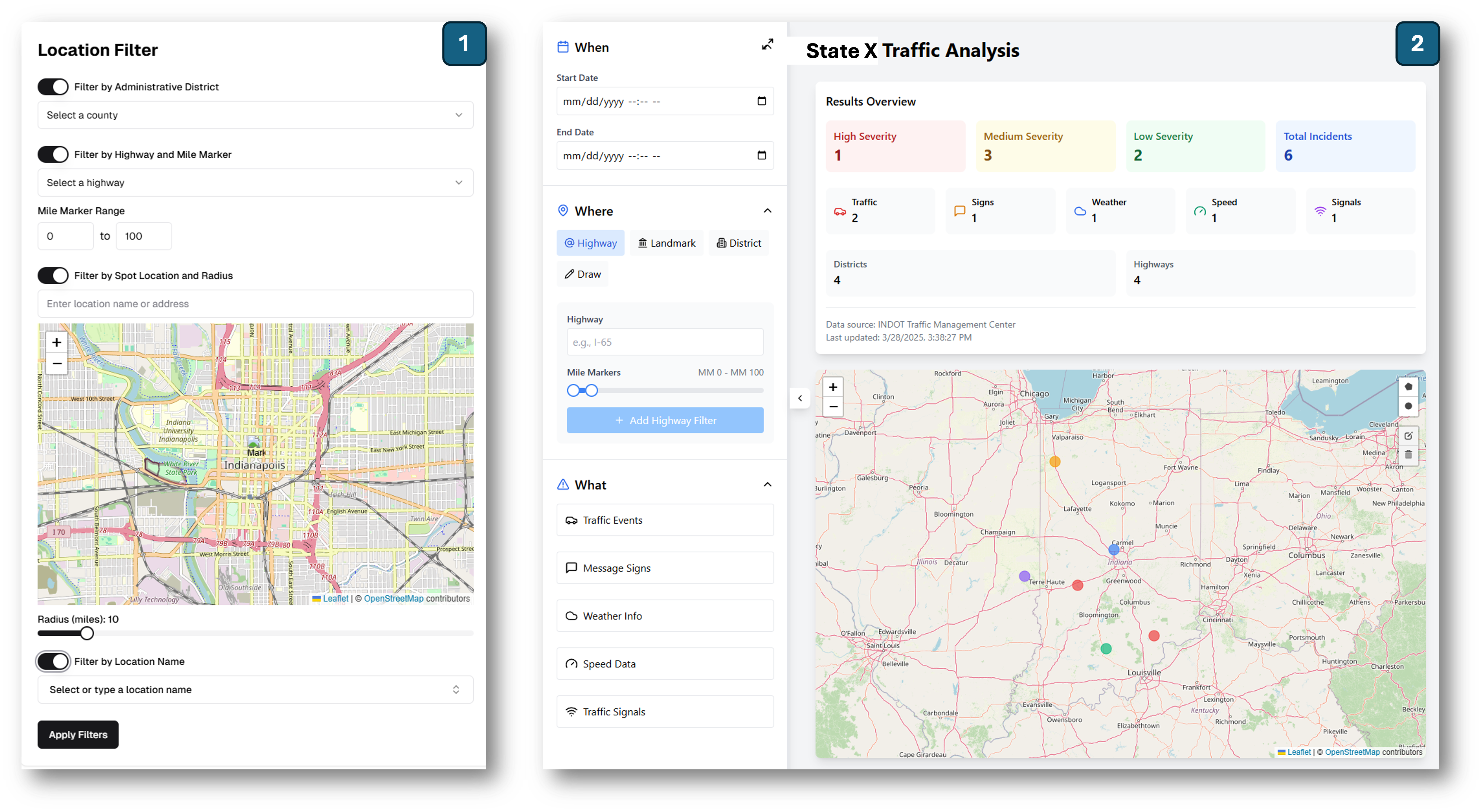}
  \caption{Example prototype applications created with generative UI. (1) focuses on brainstorming designs for specific components (created by v0) and (2) focuses on the entire page layout (created by Bolt.new).}
  \Description{}
  \label{fig:teaser}
\end{teaserfigure}


\maketitle



\section{Introduction and Background}
User-Centered Design (UCD) emphasizes early and active user involvement, coupled with frequent prototype evaluations~\cite{gould1985designing}. Applying UCD principles in practice typically involves iterative cycles of interviews and design probes~\cite{10.1145/2470654.2466473}, particularly when developing user interfaces tailored to specific domain tasks~\cite{10.1145/564376.564455, 10.1145/3613904.3642563}.

\begin{figure}[htbp]
    \centering
    \includegraphics[width=0.9\linewidth]{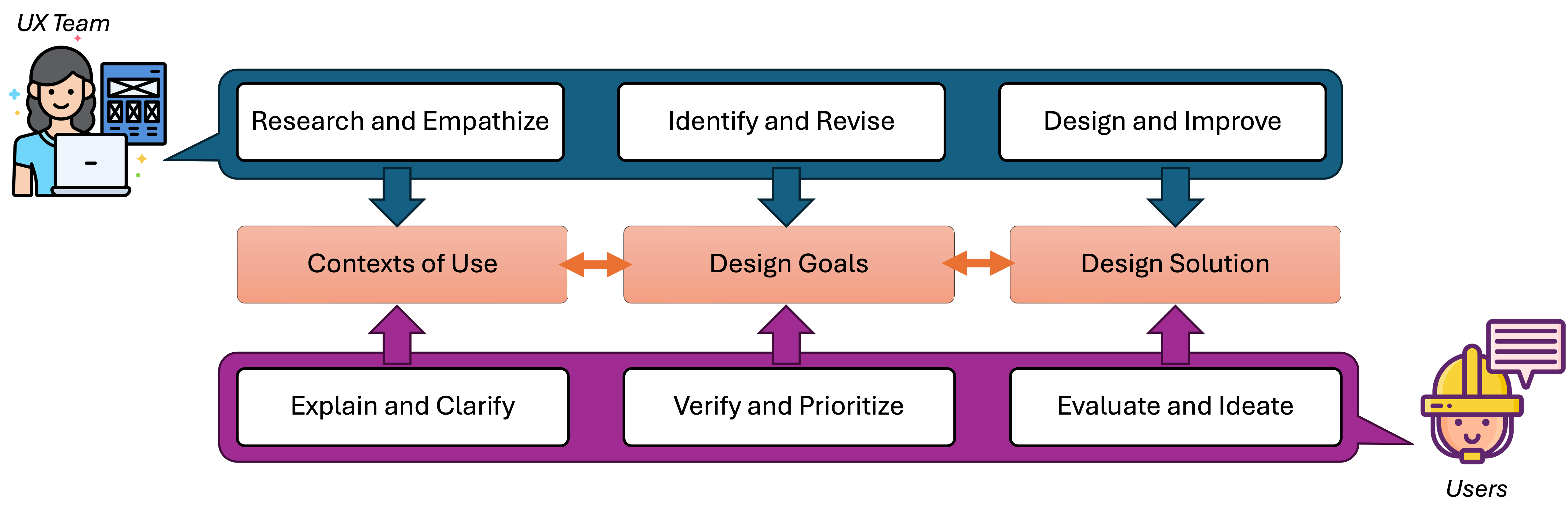}
    \caption{User-Centered Design process modeled based on the key deliverables to capture the interaction and communication between the UX team and domain expert users.}
    \label{fig:traditional-UCD}
\end{figure}

Don Norman~\cite{10.5555/2187809} describes the human-centered design process as consisting of four iterative activities: Observation, Idea Generation, Prototyping, and Testing. 
Rather than framing with UCD process with the four stages from the UX team perspective, we describe how UX team and domain-experts collaborate within the classic UCD process through the key intermediate deliverables: ``context of use,'' ``design goals,'' and ``design solutions.'' (~\autoref{fig:traditional-UCD} ) Within this iterative framework, the User Experience (UX) team initiates the drafting of each deliverable, and domain-experts actively engage in evaluating and providing feedback on these drafts. Subsequently, the UX team refines and revises the deliverables based on user feedback. 

Communication between designers and domain experts evolves across the above-mentioned stages. Initially, interactions rely heavily on interviews and surveys to gather the needs and requirements from domain experts. As the design process advances, communication increasingly leverages visual artifacts like sketches and low-fidelity (lo-fi) prototypes~\cite{reeves1992supporting, 10.1007/978-3-642-23774-4_17}. Rapid prototyping, starting with simple sketches and progressing into wireframes, mockups and high fidelity prototypes, helps clarify use contexts, user flows, and overall design goals~\cite{WILSON1988859, 10.1145/3301275.3302294}.

Early-stage prototypes, such as sketches and lo-fi wireframes, often require Wizard-of-Oz techniques~\cite{10.1145/1028014.1028086}, where designers manually demonstrate interactions because static visuals cannot fully convey intended behaviors. This frequently causes misunderstandings, resulting in additional iterations~\cite{shinohara2020design, 10.1145/3654777.3676408}. Parallel prototyping, that is, simultaneously creating multiple variations, promotes divergent thinking and collaborative solution discovery~\cite{10.1145/1879831.1879836}, yet developing prototypes for data-intensive analytics interfaces typically demands significant time and effort~\cite{kleppmann2019designing}.


\section{A Revised Ideate-Prototype Process with Generative AI in the Loop}
Recent advances in AI-assisted coding have introduced a novel paradigm coined as ``vibe coding''~\cite{karpathy2025vibe}, in which users can leverage large language models (LLMs) to generate code directly from natural-language descriptions. This approach has a great potential to shorten the path from design goals to possible design solutions, by significantly reducing the efforts needed to ideate new designs and to create functional prototypes~\cite{subramonyam2024content}. 

We introduce an initial flow for an ideation-prototyping process enabling UX teams to rapidly create high-fidelity prototypes as front-end applications using natural language prompts. Unlike the traditional sequence of sketches, wireframes, mock-ups, and prototypes, the proposed process begins directly with high-fidelity prototypes generated from user-defined design concepts and goals through AI-assisted front-end development.
Acknowledging inherent limitations of AI-generated code such as restricted complexity and potential bugs, this framework complements, rather than replaces, the low-fidelity wireframes within the iterative design process. The AI agent serves two primary roles: (1) generating design ideas based on UX designers' specified goals, ranging from overall layouts to specific component interactions, and (2) transforming low-fidelity visual designs provided by UX designers into interactive prototypes.
By shortening the journey from conceptualization to interactive prototypes, this framework aims to facilitate communication between UX teams and domain-experts with more ready-for-interaction design artifacts. 
\begin{figure}
    \centering
    \includegraphics[width=0.9\linewidth]{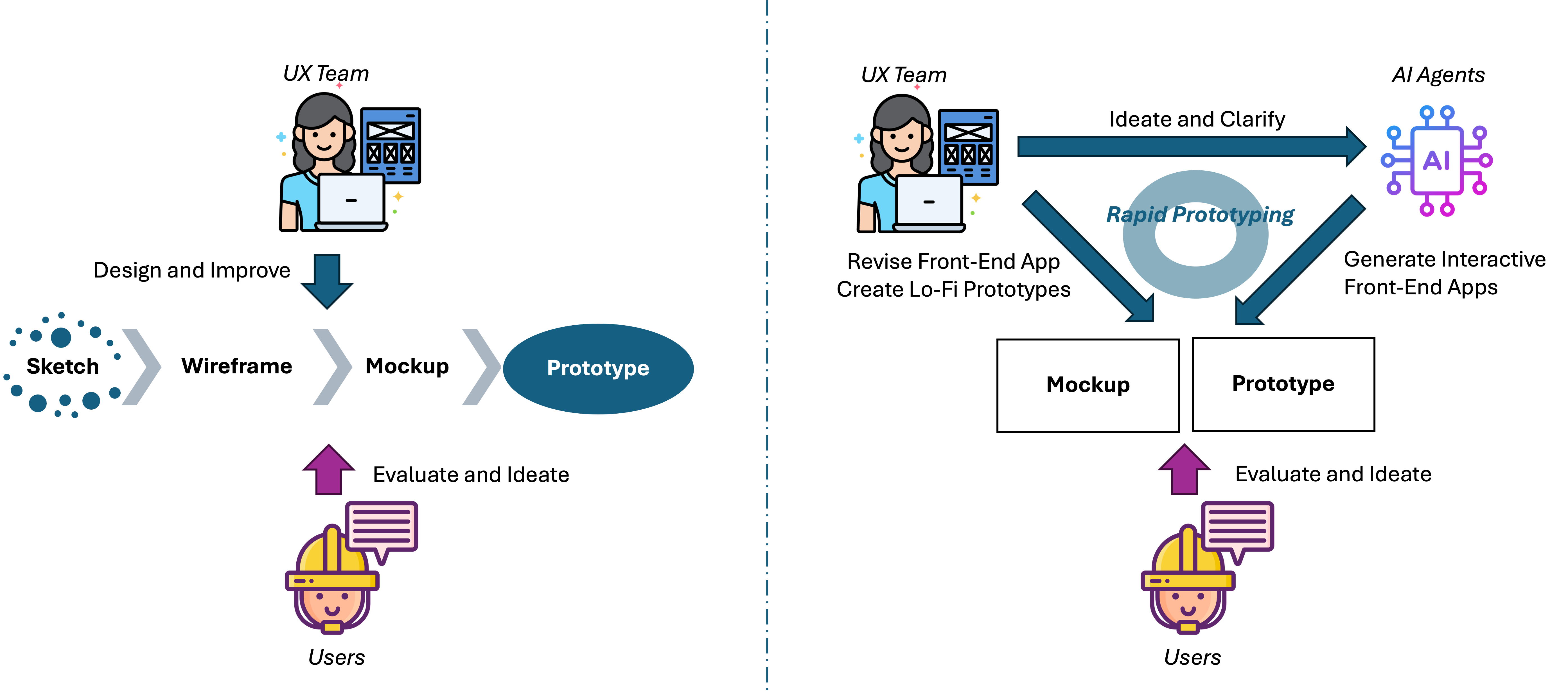}
    \caption{Comparison of Prototyping Process with (right) and without (left) Generative User Interface}
    \label{fig:new-UCD-process}
\end{figure}
\section{Case Study: Leveraging Vibe Coding in User-Centered Prototyping}

\subsection{Project Context and Objectives}
In the United States, transportation information for the public is commonly provided through the three-digit number 5-1-1, widely adopted by state and provincial transportation departments.
This project develops an intelligent user interface to efficiently access and analyze real-time 511 data from Indiana. The Indiana 511 data, includes detailed (1) traffic events, (2) live road conditions, and (3) weather-related metrics impacting road safety. 

The intended users of this interface are traffic engineers who monitor highway conditions, identify causes of slowdowns, and coordinate solutions. Although engineers currently have internal tools, they lack support for storing and analyzing historical 511 data, which this project will enable them to do.

The project team includes a senior researcher with eight years of experience in user-centered design, a graduate student skilled in cloud computing and web development, and an undergraduate student experienced in wireframe design, web programming, and information technology. All team members serve the triadic roles of designers, developers, and researchers in the user-centered design process.

\subsection{User-Centered Design Process}
We followed the four-stage human-centered design process outlined by Don Norman~\cite{10.5555/2187809}. 
Initially, the team investigated the 511 data, traffic terminology, and publicly available resources from the Indiana Department of Transportation, while also reviewing existing benchmark tools to identify potential data analysis needs.

Next, we conducted user interviews to gain insights into specific analysis requirements, workflows, and contextual factors. These interviews clarified domain-specific concepts, highlighted critical dataset features, and helped us understand relationships between data subsets relevant to analytical tasks.

Based on the interview outcomes, the team formulated a set of design goals. In this case study, we concentrate on one design goal related to retrieving relevant traffic data: to help users efficiently navigate the extensive array of data attributes and identify fields related to their query requirements. This goal addresses the complexity of the 511 data structure. For instance, each traffic event comprises 27 attributes, several containing nested sub-attributes. One illustrative example is the attribute "positiveLaneBlockage," which itself includes six sub-attributes, such as "entranceRampAffected" and "insideShoulderAffected." 

The project team compared two ideate-prototyping processes. The first followed the classic UCD process and the second leveraged vibe coding. 

\subsection{Applying the Classic UCD Process}
The team first conducted a design workshop where each member sketched their design concepts using pen and paper, then shared and discussed their ideas. The team examined the strengths and weaknesses of each design and brainstormed additional enhancements. 
\begin{figure}
    \centering
    \includegraphics[width=\linewidth]{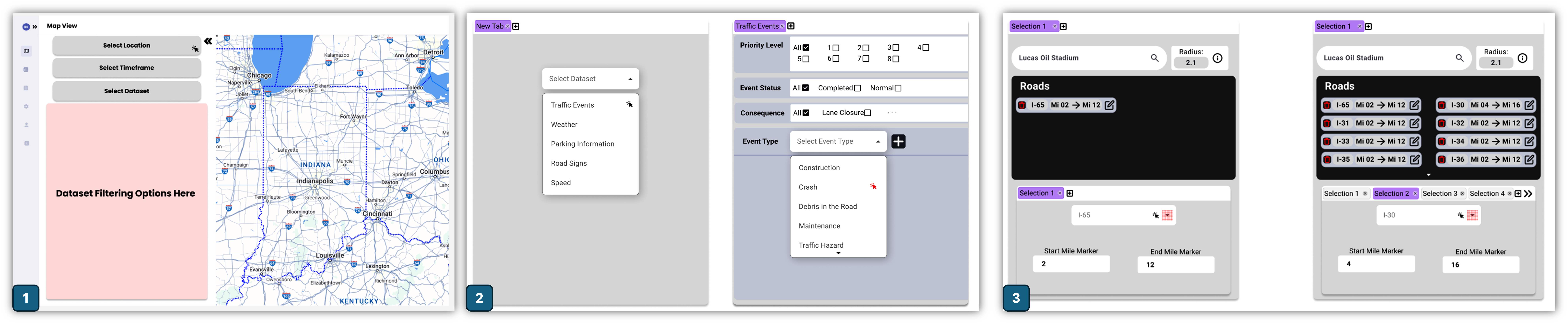}
    \caption{Example of the low-fidelity prototypes created by the team. Left (1): Example page of the overall page layout. Middle (2): Example pages for filter design to select traffic event attributes. Right (3): Example pages for filter design to define a location scope.}
    \label{fig:figma}
\end{figure}
The team then created various low-fidelity designs in Figma\footnote{\url{https://www.figma.com}}, including an overall page layout with multiple interaction variations, as well as components for dataset, time, and location selection, each featuring numerous interaction options. Additionally, widgets were developed to support task-specific filters, such as selecting recurring dates for common scenarios like weekday commutes.


\subsection{Ideate and Prototyping with Vibe Coding}
The team used two generative UI tools V0~\footnote{\url{https://v0.dev/} Developed by Vercel, an AI coding assistant that generates user interfaces and code from plain English prompts.} and Bolt.new~\footnote{\url{https://bolt.new/} Integrates AI with WebContainers, allows developers to create, run, debug, and instantly deploy full-stack applications in-browser, using natural language prompts} to synthesize existing ideas and brainstorm alternative designs. 
Both platforms support a range of web technologies and frameworks, such as React. We used both platforms' free plans in our design process.

\paragraph{Brainstorming Alternative Designs}
One challenge when creating low-fidelity prototypes is that static interface designs do not fully convey the nuances of data-intensive applications effectively. The primary difficulties lie less in the visual design of UI widgets and more in effectively presenting and visualizing data attributes and values to facilitate interactive sensemaking and data query.

We experimented with providing example JSON objects in prompts and requesting potential interfaces for data filtering. Due to chat context window limitations in both tools, we included conceptual JSON structures without actual data as attachments. Our prompts included specific contexts such as: ``Please create a web interface that allows users to retrieve historical data using the attached schema. Data has been collected every minute over five years, accompanied by a design goal clearly stating, ``The design goal is to help users efficiently navigate the extensive array of data attributes and identify fields related to their query requirements.'' We also added ``Please generate dummy data for the provided schema to illustrate the interface functionality.''

These prompting strategies produced diverse layout alternatives, such as a side-by-side arrangement with filter options displayed on the left or a tabbed view to switch between search criteria and results. These AI-generated design inspired the team to refine the overall interface layout, clarifying the conceptual organization of features. Additionally, it allowed the team to ``zoom out,'' reminding the team of general features overlooked when deeply focused on domain-specific requirements. Examples include adding a ``clear'' button to reset fields and providing options to export results in tabular formats.

Furthermore, the AI-generated designs introduced innovative interaction patterns previously unexplored by the team. Initially, our design included static attribute filters requiring users to specify value ranges upfront. In contrast, one AI-generated interface suggested a modular approach, enabling users to dynamically create filters by selecting fields, operators, and value ranges as needed.

For detailed feature refinement, we experimented with iterative prompting within the same session, e.g., ``The user prefers to define locations by administrative districts, highway names with mile markers, and location names with radius range. Update the location selection accordingly.'' However, this approach led to cumulative errors and integration challenges after several iterations. Resetting the conversation in a new session also posed consistent integration challenges for the various designs produced.

\paragraph{Generating Interactive Versions of Existing Designs}
V0 and Bolt support connections to Figma; however, this integration requires Figma files to be organized with each design on a separate page. As the team engaged in rapid prototyping, multiple design iterations were created as frames within a single page. Thus, direct imports would require additional efforts to organize the design pages. We uploaded screenshots of individual frames to generate code implementation of the visual designs. The results transformed the design accurately with static web pages, but did not correctly implement the desired interactions.

\paragraph{Synthesize Existing Designs}
The team also experimented by prompting both tools to create synthesized designs based on insights from the first two interview rounds. We provided detailed descriptions of the improved design, including envisioned layouts, supported interactions, and key functionalities (see \autoref{sec:appendix}). Notably, both tools featured an ``Enhance Prompt'' option that rewrites user-provided prompts to be more comprehensive and context-aware. This enhancement mechanism was used before generating the user interfaces. The examples provided in the Appendix are raw prompts before applying this enhancement.
Overall, using generative UI at this stage resulted in the effective design that prepares the team for the implementation stages.

\section{Discussion}
In this case study, we applied both a classical user-centered design (UCD) process and an AI-assisted UCD process to design and develop a data analytics interface for traffic engineers. Below, we reflect on the advantages and challenges associated with using vibe coding in user-centered design.

\subsection{Scaffolding Ideation and the Design Process}

The design team initially spent three months understanding the context of use, including exploring the 511 dataset and domain-specific terminology. Given that the 511 data is publicly available, popular LLM services such as ChatGPT and Gemini, and generative UI tools like V0 and Bolt, already had prior knowledge about the dataset that surpassed the team's initial familiarity. If the team had leveraged generative AI earlier in creating user interfaces, it could have greatly accelerated their ideation process.

Additionally, several team members had limited experience with wireframing and mockup design tools like Figma. Creating low-fidelity designs was time-consuming and mentally taxing. This challenge sometimes constrained the ability to adjust design details effectively and hindered broader strategic thinking. Utilizing generative UI tools allowed team members to dedicate more mental resources to strategic evaluation and iterative improvement of generated designs. This facilitated the production of a larger number of higher quality designs according to the goals and requirements gathered from domain experts.

\subsection{Accelerating Prototyping and User Feedback}

Employing generative UI significantly reduced the time required to communicate design ideas among team members and with the users. Although frequently required refinement, these UI outputs could serve as effective static designs or be improved through iterative prompting.

Furthermore, providing more interactive and realistic prototypes beyond mere conceptual sketches and illustrations improved the quality and quantity of feedback gathered from users. When initially presented with sketches and low-fidelity designs, users tended to refrain from detailed critiques, considering them preliminary representations rather than concrete implementations. Consequently, early-stage design discussions often failed to capture detailed user feedback and missed opportunities to uncover additional requirements. In contrast, interactive prototypes accessible via shared links allowed users to directly engage with the design. This tangible interaction encouraged users to actively test the prototypes, uncover potential issues, and collaboratively ideate valuable additional features and refinements.

\subsection{Potential Risks and Ethical Considerations}
Ultimately, synthesizing AI-generated designs into functional applications demands rigorous programming practices and structured workflows. 
While generative UI offers substantial benefits, it also introduces certain challenges and risks. The AI-generated code occasionally contains errors requiring extensive debugging conversations, which deviates from the conversation about design ideas. Additionally, there are increasing concerns regarding the security of AI-generated code~\cite{9833571}. In this study, we leveraged vibe coding primarily for design probes and interactive prototyping; however, incorporating AI-generated code directly into development demands careful consideration and rigorous testing. 

A notable risk of employing generative UI in ideation is the potential to unintentionally converge on conventional design patterns, potentially hindering innovative thinking. Effective implementation requires carefully selecting the timing for introducing AI assistance, according to the team expertise and project priorities. For example, for teams with experienced and creative designers, AI assistance may be more useful for translating static designs into interactive code, rather than as an ideation support. Yet, for novice designers or teams with limited design experience, generative UI can help bridge this expertise gap. Moreover, successfully leveraging generative AI necessitates proficiency in prompt engineering and deep domain knowledge. Thus, when integrating AI into UCD, designers' expertise and AI literacy must be carefully considered.  

Utilizing vibe coding within user-centered design highlights several critical considerations for responsible AI use. While vibe coding greatly expands creative possibilities and expedites the design cycle, it also underscores the necessity for clear, detailed instructions to avoid misalignment between generated designs and user expectations. Designers also need to be careful with what information and data to share with the AI agent when generating designs. 

\begin{acks}
This work was supported in part by the Joint Transportation Research Program administered by the Indiana Department of Transportation and Purdue University.   The contents of this manuscript reflect the views of the authors, who are responsible for the facts and the accuracy of the data presented herein. The contents do not necessarily reflect the official views and policies of the Indiana Department of Transportation or the Federal Highway Administration. The report does not constitute a standard, specification or regulation.
\end{acks}

\bibliographystyle{ACM-Reference-Format}
\bibliography{acmart}


\begin{thebibliography}{17}


\ifx \showCODEN    \undefined \def \showCODEN     #1{\unskip}     \fi
\ifx \showISBNx    \undefined \def \showISBNx     #1{\unskip}     \fi
\ifx \showISBNxiii \undefined \def \showISBNxiii  #1{\unskip}     \fi
\ifx \showISSN     \undefined \def \showISSN      #1{\unskip}     \fi
\ifx \showLCCN     \undefined \def \showLCCN      #1{\unskip}     \fi
\ifx \shownote     \undefined \def \shownote      #1{#1}          \fi
\ifx \showarticletitle \undefined \def \showarticletitle #1{#1}   \fi
\ifx \showURL      \undefined \def \showURL       {\relax}        \fi
\providecommand\bibfield[2]{#2}
\providecommand\bibinfo[2]{#2}
\providecommand\natexlab[1]{#1}
\providecommand\showeprint[2][]{arXiv:#2}

\bibitem[Dow et~al\mbox{.}(2011)]%
        {10.1145/1879831.1879836}
\bibfield{author}{\bibinfo{person}{Steven~P. Dow}, \bibinfo{person}{Alana Glassco}, \bibinfo{person}{Jonathan Kass}, \bibinfo{person}{Melissa Schwarz}, \bibinfo{person}{Daniel~L. Schwartz}, {and} \bibinfo{person}{Scott~R. Klemmer}.} \bibinfo{year}{2011}\natexlab{}.
\newblock \showarticletitle{Parallel prototyping leads to better design results, more divergence, and increased self-efficacy}.
\newblock \bibinfo{journal}{\emph{ACM Trans. Comput.-Hum. Interact.}} \bibinfo{volume}{17}, \bibinfo{number}{4}, Article \bibinfo{articleno}{18} (\bibinfo{date}{Dec.} \bibinfo{year}{2011}), \bibinfo{numpages}{24}~pages.
\newblock
\showISSN{1073-0516}
\href{https://doi.org/10.1145/1879831.1879836}{doi:\nolinkurl{10.1145/1879831.1879836}}


\bibitem[Gould and Lewis(1985)]%
        {gould1985designing}
\bibfield{author}{\bibinfo{person}{John~D Gould} {and} \bibinfo{person}{Clayton Lewis}.} \bibinfo{year}{1985}\natexlab{}.
\newblock \showarticletitle{Designing for usability: key principles and what designers think}.
\newblock \bibinfo{journal}{\emph{Commun. ACM}} \bibinfo{volume}{28}, \bibinfo{number}{3} (\bibinfo{year}{1985}), \bibinfo{pages}{300--311}.
\newblock


\bibitem[Hansen et~al\mbox{.}(2002)]%
        {10.1145/564376.564455}
\bibfield{author}{\bibinfo{person}{Preben Hansen}, \bibinfo{person}{Daniela Petrelli}, \bibinfo{person}{Jussi Karlgren}, \bibinfo{person}{Micheline Beaulieu}, {and} \bibinfo{person}{Mark Sanderson}.} \bibinfo{year}{2002}\natexlab{}.
\newblock \showarticletitle{User-centered interface design for cross-language information retrieval}. In \bibinfo{booktitle}{\emph{Proceedings of the 25th Annual International ACM SIGIR Conference on Research and Development in Information Retrieval}} (Tampere, Finland) \emph{(\bibinfo{series}{SIGIR '02})}. \bibinfo{publisher}{Association for Computing Machinery}, \bibinfo{address}{New York, NY, USA}, \bibinfo{pages}{383–384}.
\newblock
\showISBNx{1581135610}
\href{https://doi.org/10.1145/564376.564455}{doi:\nolinkurl{10.1145/564376.564455}}


\bibitem[Karpathy(2025)]%
        {karpathy2025vibe}
\bibfield{author}{\bibinfo{person}{Andrej Karpathy}.} \bibinfo{year}{2025}\natexlab{}.
\newblock \bibinfo{title}{There's a new kind of coding I call "vibe coding", where you fully give in to the vibes, embrace exponentials, and forget that the code even exists.}
\newblock
\urldef\tempurl%
\url{https://x.com/karpathy/status/1886192184808149383}
\showURL{%
\tempurl}


\bibitem[Kleppmann(2019)]%
        {kleppmann2019designing}
\bibfield{author}{\bibinfo{person}{Martin Kleppmann}.} \bibinfo{year}{2019}\natexlab{}.
\newblock \bibinfo{title}{Designing data-intensive applications}.
\newblock


\bibitem[Li et~al\mbox{.}(2019)]%
        {10.1145/3301275.3302294}
\bibfield{author}{\bibinfo{person}{Tianyi Li}, \bibinfo{person}{Gregorio Convertino}, \bibinfo{person}{Ranjeet~Kumar Tayi}, {and} \bibinfo{person}{Shima Kazerooni}.} \bibinfo{year}{2019}\natexlab{}.
\newblock \showarticletitle{What data should I protect? recommender and planning support for data security analysts}. In \bibinfo{booktitle}{\emph{Proceedings of the 24th International Conference on Intelligent User Interfaces}} (Marina del Ray, California) \emph{(\bibinfo{series}{IUI '19})}. \bibinfo{publisher}{Association for Computing Machinery}, \bibinfo{address}{New York, NY, USA}, \bibinfo{pages}{286–297}.
\newblock
\showISBNx{9781450362726}
\href{https://doi.org/10.1145/3301275.3302294}{doi:\nolinkurl{10.1145/3301275.3302294}}


\bibitem[Molin(2004)]%
        {10.1145/1028014.1028086}
\bibfield{author}{\bibinfo{person}{Lennart Molin}.} \bibinfo{year}{2004}\natexlab{}.
\newblock \showarticletitle{Wizard-of-Oz prototyping for co-operative interaction design of graphical user interfaces}. In \bibinfo{booktitle}{\emph{Proceedings of the Third Nordic Conference on Human-Computer Interaction}} (Tampere, Finland) \emph{(\bibinfo{series}{NordiCHI '04})}. \bibinfo{publisher}{Association for Computing Machinery}, \bibinfo{address}{New York, NY, USA}, \bibinfo{pages}{425–428}.
\newblock
\showISBNx{1581138571}
\href{https://doi.org/10.1145/1028014.1028086}{doi:\nolinkurl{10.1145/1028014.1028086}}


\bibitem[Norman(2002)]%
        {10.5555/2187809}
\bibfield{author}{\bibinfo{person}{Donald~A. Norman}.} \bibinfo{year}{2002}\natexlab{}.
\newblock \bibinfo{booktitle}{\emph{The Design of Everyday Things}}.
\newblock \bibinfo{publisher}{Basic Books, Inc.}, \bibinfo{address}{USA}.
\newblock
\showISBNx{9780465067107}


\bibitem[Pearce et~al\mbox{.}(2022)]%
        {9833571}
\bibfield{author}{\bibinfo{person}{Hammond Pearce}, \bibinfo{person}{Baleegh Ahmad}, \bibinfo{person}{Benjamin Tan}, \bibinfo{person}{Brendan Dolan-Gavitt}, {and} \bibinfo{person}{Ramesh Karri}.} \bibinfo{year}{2022}\natexlab{}.
\newblock \showarticletitle{Asleep at the Keyboard? Assessing the Security of GitHub Copilot’s Code Contributions}. In \bibinfo{booktitle}{\emph{2022 IEEE Symposium on Security and Privacy (SP)}}. \bibinfo{pages}{754--768}.
\newblock
\href{https://doi.org/10.1109/SP46214.2022.9833571}{doi:\nolinkurl{10.1109/SP46214.2022.9833571}}


\bibitem[Reeves and Shipman(1992)]%
        {reeves1992supporting}
\bibfield{author}{\bibinfo{person}{Brent Reeves} {and} \bibinfo{person}{Frank Shipman}.} \bibinfo{year}{1992}\natexlab{}.
\newblock \showarticletitle{Supporting communication between designers with artifact-centered evolving information spaces}. In \bibinfo{booktitle}{\emph{Proceedings of the 1992 ACM conference on Computer-supported cooperative work}}. \bibinfo{pages}{394--401}.
\newblock


\bibitem[Sharmin and Bailey(2011)]%
        {10.1007/978-3-642-23774-4_17}
\bibfield{author}{\bibinfo{person}{Moushumi Sharmin} {and} \bibinfo{person}{Brian~P. Bailey}.} \bibinfo{year}{2011}\natexlab{}.
\newblock \showarticletitle{Making Sense of Communication Associated with Artifacts during Early Design Activity}. In \bibinfo{booktitle}{\emph{Human-Computer Interaction -- INTERACT 2011}}, \bibfield{editor}{\bibinfo{person}{Pedro Campos}, \bibinfo{person}{Nicholas Graham}, \bibinfo{person}{Joaquim Jorge}, \bibinfo{person}{Nuno Nunes}, \bibinfo{person}{Philippe Palanque}, {and} \bibinfo{person}{Marco Winckler}} (Eds.). \bibinfo{publisher}{Springer Berlin Heidelberg}, \bibinfo{address}{Berlin, Heidelberg}, \bibinfo{pages}{181--198}.
\newblock
\showISBNx{978-3-642-23774-4}


\bibitem[Shinohara et~al\mbox{.}(2020)]%
        {shinohara2020design}
\bibfield{author}{\bibinfo{person}{Kristen Shinohara}, \bibinfo{person}{Nayeri Jacobo}, \bibinfo{person}{Wanda Pratt}, {and} \bibinfo{person}{Jacob~O Wobbrock}.} \bibinfo{year}{2020}\natexlab{}.
\newblock \showarticletitle{Design for social accessibility method cards: Engaging users and reflecting on social scenarios for accessible design}.
\newblock \bibinfo{journal}{\emph{ACM Transactions on Accessible Computing (TACCESS)}} \bibinfo{volume}{12}, \bibinfo{number}{4} (\bibinfo{year}{2020}), \bibinfo{pages}{1--33}.
\newblock


\bibitem[Subramonyam et~al\mbox{.}(2024)]%
        {subramonyam2024content}
\bibfield{author}{\bibinfo{person}{Hari Subramonyam}, \bibinfo{person}{Divy Thakkar}, \bibinfo{person}{J{\"u}rgen Dieber}, {and} \bibinfo{person}{Anoop Sinha}.} \bibinfo{year}{2024}\natexlab{}.
\newblock \showarticletitle{Content-Centric Prototyping of Generative AI Applications: Emerging Approaches and Challenges in Collaborative Software Teams}.
\newblock \bibinfo{journal}{\emph{arXiv preprint arXiv:2402.17721}} (\bibinfo{year}{2024}).
\newblock
\urldef\tempurl%
\url{https://arxiv.org/abs/2402.17721}
\showURL{%
\tempurl}


\bibitem[Wallace et~al\mbox{.}(2013)]%
        {10.1145/2470654.2466473}
\bibfield{author}{\bibinfo{person}{Jayne Wallace}, \bibinfo{person}{John McCarthy}, \bibinfo{person}{Peter~C. Wright}, {and} \bibinfo{person}{Patrick Olivier}.} \bibinfo{year}{2013}\natexlab{}.
\newblock \showarticletitle{Making design probes work}. In \bibinfo{booktitle}{\emph{Proceedings of the SIGCHI Conference on Human Factors in Computing Systems}} (Paris, France) \emph{(\bibinfo{series}{CHI '13})}. \bibinfo{publisher}{Association for Computing Machinery}, \bibinfo{address}{New York, NY, USA}, \bibinfo{pages}{3441–3450}.
\newblock
\showISBNx{9781450318990}
\href{https://doi.org/10.1145/2470654.2466473}{doi:\nolinkurl{10.1145/2470654.2466473}}


\bibitem[Weitz et~al\mbox{.}(2024)]%
        {10.1145/3613904.3642563}
\bibfield{author}{\bibinfo{person}{Katharina Weitz}, \bibinfo{person}{Ruben Schlagowski}, \bibinfo{person}{Elisabeth Andr\'{e}}, \bibinfo{person}{Maris M\"{a}nniste}, {and} \bibinfo{person}{Ceenu George}.} \bibinfo{year}{2024}\natexlab{}.
\newblock \showarticletitle{Explaining It Your Way - Findings from a Co-Creative Design Workshop on Designing XAI Applications with AI End-Users from the Public Sector}. In \bibinfo{booktitle}{\emph{Proceedings of the 2024 CHI Conference on Human Factors in Computing Systems}} (Honolulu, HI, USA) \emph{(\bibinfo{series}{CHI '24})}. \bibinfo{publisher}{Association for Computing Machinery}, \bibinfo{address}{New York, NY, USA}, Article \bibinfo{articleno}{745}, \bibinfo{numpages}{14}~pages.
\newblock
\showISBNx{9798400703300}
\href{https://doi.org/10.1145/3613904.3642563}{doi:\nolinkurl{10.1145/3613904.3642563}}


\bibitem[Wilson and Rosenberg(1988)]%
        {WILSON1988859}
\bibfield{author}{\bibinfo{person}{James Wilson} {and} \bibinfo{person}{Daniel Rosenberg}.} \bibinfo{year}{1988}\natexlab{}.
\newblock \showarticletitle{Chapter 39 - Rapid Prototyping for User Interface Design}.
\newblock In \bibinfo{booktitle}{\emph{Handbook of Human-Computer Interaction}}, \bibfield{editor}{\bibinfo{person}{MARTIN HELANDER}} (Ed.). \bibinfo{publisher}{North-Holland}, \bibinfo{address}{Amsterdam}, \bibinfo{pages}{859--875}.
\newblock
\showISBNx{978-0-444-70536-5}
\href{https://doi.org/10.1016/B978-0-444-70536-5.50044-0}{doi:\nolinkurl{10.1016/B978-0-444-70536-5.50044-0}}


\bibitem[Wu et~al\mbox{.}(2024)]%
        {10.1145/3654777.3676408}
\bibfield{author}{\bibinfo{person}{Jason Wu}, \bibinfo{person}{Yi-Hao Peng}, \bibinfo{person}{Xin Yue~Amanda Li}, \bibinfo{person}{Amanda Swearngin}, \bibinfo{person}{Jeffrey~P Bigham}, {and} \bibinfo{person}{Jeffrey Nichols}.} \bibinfo{year}{2024}\natexlab{}.
\newblock \showarticletitle{UIClip: A Data-driven Model for Assessing User Interface Design}. In \bibinfo{booktitle}{\emph{Proceedings of the 37th Annual ACM Symposium on User Interface Software and Technology}} (Pittsburgh, PA, USA) \emph{(\bibinfo{series}{UIST '24})}. \bibinfo{publisher}{Association for Computing Machinery}, \bibinfo{address}{New York, NY, USA}, Article \bibinfo{articleno}{45}, \bibinfo{numpages}{16}~pages.
\newblock
\showISBNx{9798400706288}
\href{https://doi.org/10.1145/3654777.3676408}{doi:\nolinkurl{10.1145/3654777.3676408}}


\end{thebibliography}

\appendix
\section{Example (Raw) Prompts}
\label{sec:appendix}
Below are two example prompts that we have provided to both tools to synthesize the designs and insights from the user studies.

For generating the overall layout:
\begin{quote}
    Build a React app with a collapsible side menu on the left. It has five options: Map View, Chart View, Data Catalog, Settings, and Guide. A toggle button switches between 5\% and 15\% width.
To the right, divide the screen into a map area and a selector panel. The map uses Leaflet.js with checkboxes for "Dynamic Message Signs" and "Traffic Timing System." It takes up two-thirds of the space.
The selector panel includes buttons: Select Location, Select Timeframe, Select Dataset, Save Query, and Fetch New Data. Each main button opens a collapsible card with filters. The panel is resizable and collapsible, showing a summary when closed.
The Location card has four tabs: Road, Named Location, Districts, and Draw. Each tab offers different ways to select locations, using dropdowns, range sliders, or drawing tools.
Timeframe selection supports date ranges, time ranges, recurring selections, and holidays. All values show in ISO format. The calendar should fully overlay the map.
The Dataset card has five tabs: Car Events, Lane Blockage, Rest Area, Social Events, and Weather. Each has filter forms, some with dropdowns or checkboxes. Apply Filter shows chosen values at the top. Social and Weather tabs can use placeholders.
Use collapsible layouts and stacked tabs to avoid clutter. “Select All” toggles should work consistently across dropdowns. Show filter selections clearly without overwhelming the UI.
\end{quote}
\begin{figure}[htbp]
    \centering
    \begin{minipage}[b]{0.48\textwidth}
        \centering
        \includegraphics[width=\textwidth]{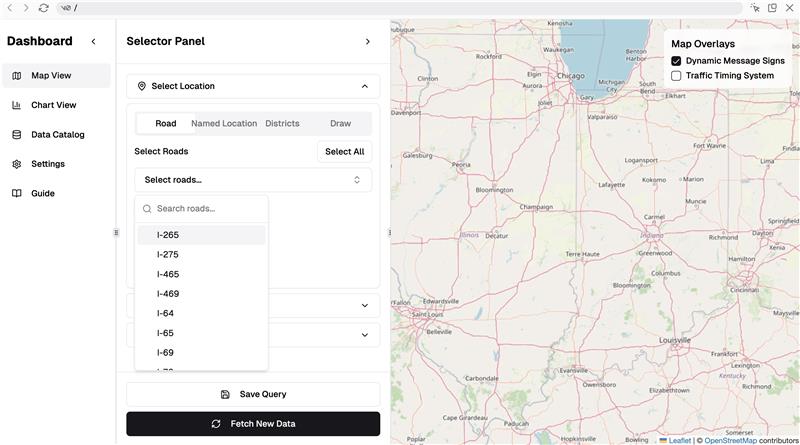}
        \caption{Results from running the prompt for overall layout with Bolt.new.}
        \label{fig:bolt-1}
    \end{minipage}
    \hfill
    \begin{minipage}[b]{0.48\textwidth}
        \centering
        \includegraphics[width=\textwidth]{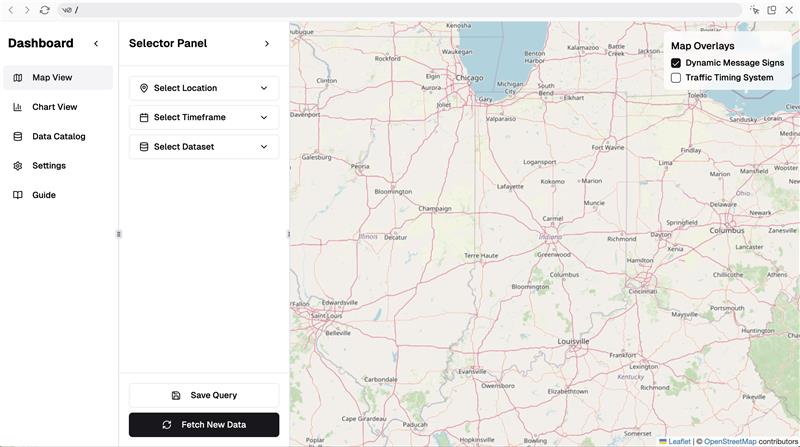}
        \caption{Results from running the prompt for overall layout with v0.}
        \label{fig:v0-1}
    \end{minipage}
\end{figure}

For generating additional features:
\begin{quote}
    Develop a React-based user interface (UI) to visualize and interact with traffic data. The UI should feature a split-screen layout: the left side should contain a panel for filtering traffic data, and the right side should display a Leaflet map. The map should dynamically update based on the filter selections. Implement a time slider control within the map view to enable users to explore how traffic data changes over time. Include functionality to save the current filter selections, allowing users to load and restore these selections later. Below the map, incorporate a collapsible results card that summarizes the filtered data in a concise and easily understandable format. Ensure the UI is responsive, providing a seamless experience across various screen sizes, and maintain a clean, intuitive design for ease of navigation.
\end{quote}
\begin{figure}[htbp]
    \centering
    \begin{minipage}[b]{0.48\textwidth}
        \centering
        \includegraphics[width=\textwidth]{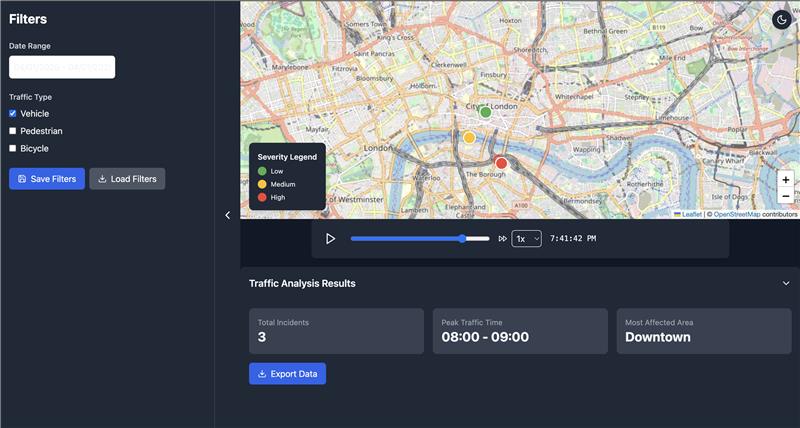}
        \caption{Results from running the prompt for additional features with Bolt.new.}
        \label{fig:bolt-1}
    \end{minipage}
    \hfill
    \begin{minipage}[b]{0.48\textwidth}
        \centering
        \includegraphics[width=\textwidth]{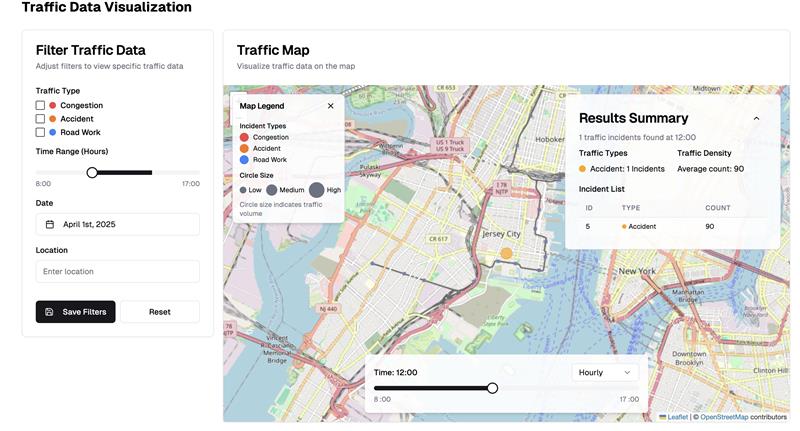}
        \caption{Results from running the prompt for additional features with v0.}
        \label{fig:v0-1}
    \end{minipage}
\end{figure}

For adding additional features:
\begin{quote}
    There should be a button at the top to move the filter panel from the right side of the screen to the left, and visa versa. Also the Save Query and Fetch Data buttons at the bottom should only be on the same row if the panel is wide enough to display all of the text of the button without overlap. Otherwise the two buttons need to be on different rows. Additionally, the "Save Query" form that pops up should be the width of the filter panel.
 
The Move filter button should be a part of the top menu of the filter panel. Also when the full screen button is clicked on the map, it should full screen only the map.
\end{quote}
\end{document}